\documentstyle[prl,epsfig,aps]{revtex}
 \newcommand {\bi} {\bibitem}
 \newcommand {\be} {\begin{equation}}
\newcommand {\bea} {\begin{eqnarray} \nonumber }
\newcommand {\ee} {\end{equation}}
\newcommand {\eea} {\end{eqnarray}}

 \newcommand {\al} {\alpha}

\newcommand {\cH}  {{\cal H}}

\newcommand {\cN}  {{\cal N}}

\newcommand {\Tr} {\mbox{Tr}}

\newcommand {\ato}  {\left(}

\newcommand {\cto}  {\right)}

\def \form#1 {eq. (\ref{#1}) }
\def \parziale#1#2  {{\partial {#1} \over \partial {#2}}}

\begin{document}
\twocolumn
[\hsize\textwidth\columnwidth\hsize\csname@twocolumnfalse\endcsname

\title{The thermodynamical
liquid-glass transition in a Lennard-Jones binary mixture}
\author{Barbara Coluzzi}\address{John von Neumann-Institut f\"ur Computing (NIC) \\
c/o Forschungszentrum J\"ulich, 
D-52425 J\"ulich (Germany)}
\author{
Giorgio
Parisi and Paolo Verrocchio}
\address{ Dipartimento di Fisica, Universit\`a {\em La  Sapienza},\\ 
INFN Sezione di Roma I, 
Piazzale Aldo Moro, Rome 00185 (Italy)}
\maketitle
\begin{abstract}
We use the results derived in the framework of the replica approach to study the liquid-glass 
thermodynamic transition.  The main results are derived without using replicas and applied to the 
study of the Lennard-Jones binary mixture introduced by Kob and Andersen. We find that there is a 
phase transition due to the entropy crisis. We compute both analytically and numerically the value 
of the phase transition point $T_{K}$ and the specific heat in the low temperature phase.

\end{abstract}
\vskip.3cm
]
\narrowtext

In recent times there have been many progresses in  the analytic understanding of the thermodynamics 
transition of glasses \cite{MePa1,Me,sferesoft,LJ}.
Beyond technicalities the basic assumption is that {\sl a glass is very near to a frozen liquid}.  
This quite old statement can be rephrased as {\sl a liquid is very near to a heated glass}.  In 
other words the configurations of a glass at low temperature are not far from those of a liquid.  
Therefore, if we use some smart method to explore the phase space in the liquid phase, we can find the 
properties of the low temperature glassy phase.

This strategy has been put into action using the replica theory.  One finds, as output 
of an explicit computation, that the glass transition is characterized by the vanishing of the 
configurational entropy (the so called complexity) and the low temperature 
phase is described by {\sl one step replica symmetry breaking} \cite{KiThWo} with a non-vanishing 
non-ergodic parameter at the phase transition point.  

Replica theory is a very powerful tool, but it has the disadvantage that many of the underlying 
physical hypothesis cannot be seen in a clear way.  We will rederive some of 
the main result of \cite{MePa1} without using the replica formalism \cite{Mo}.  We suppose 
that below some temperature (to be identified with the mode coupling transition \cite{GoSj}), the 
phase space of the system can be approximately divided into regions (which we will call valleys) which 
are separated by high barriers \cite{SaDeSt,landscape}.  This approximation becomes better and better when 
we decrease the temperature and it becomes exact below the thermodynamics phase transition 
temperature $T_{K}$ of the glass \cite{Ka}, where the viscosity should diverges \cite{An} 
(unfortunately no quantitative predictions have been obtained microscopically on the behaviour of 
the viscosity beyond the Adams Gibbs argument). In a first 
approximation each valley  can be  associated to one inherent structure, i.e. one minimum of the 
potential energy \cite{SaDeSt,St,KST1}.

Let us consider a generic system with $N$ particles with Hamiltonian $H(C)$, $C$ denoting the 
generic configuration of the system.  The partition function both in the 
liquid phase at low temperature and in the glass phase can  be written as
\bea
Z(\beta)=
\sum_{a} \exp( -\beta N f_{a}(\beta))=
\\
\int d \cN(f,\beta) \exp (-\beta N f),
\eea
where $f_{a}(\beta)$ is the free energy density of the valley 
labeled by $a$ at the temperature $\beta^{-1}$ 
and $\cN(f,\beta)$ is the number of valleys with free energy density (per particle) less than $f$.  
In the glassy phase this sum is dominated by the valleys with minimal free energy, while in the 
liquid phase an exponentially large number of valleys contributes to the partition function.  It is 
normally assumed that in glass forming systems, for large $N$ and $f>f_{0}(\beta)$, we have that
\be
\cN(f,\beta) =\exp(N \Sigma(f,\beta)),
\ee
where the configurational entropy, or complexity, $\Sigma(f,\beta)$ is positive in this region and 
vanishes at $f=f_{0}(\beta)$.  The quantity $ f_{0}(\beta)$ is the minimum value of the free energy: 
$\cN(f,\beta)$ is zero for $f< f_{0}(\beta)$.

The key point is the computation of the function $\Sigma(f,\beta)$ for $f>f_{0}(\beta)$. It can be 
done using liquid theory because in this region we are still in the liquid phase. The location of 
the zero of  $\Sigma(f,\beta)$ will tell us the value of the free energy in the glassy phase.

At this end it is convenient to consider the generalized partition function
\be
Z(\gamma;\beta)\equiv \exp ( -N \gamma \Phi(\gamma;\beta))=\sum_{a} \exp( -\gamma N f_{a}(\beta)).
\ee
The physical meaning of $\gamma$ is will be clearer later.

It is evident that 
\be
\gamma \Phi(\gamma;\beta)=  \gamma f - \Sigma(\beta,f),\ \ \ 
f={\partial (\gamma (\Phi(\gamma;\beta) ))\over \partial \gamma}.
\ee
The complexity can be simply obtained from $\Phi(\gamma;\beta)$ in the same way as the entropy 
can be obtained from the usual free energy.

The crucial step consists in writing
\bea
Z(\gamma;\beta) = \nonumber \\
\int  d C \exp \ato-\gamma H(C) -N \gamma  f(\beta,C) + N\gamma 
f(\gamma,C)\cto = \nonumber \\
\int  d C \exp \ato-\gamma H(C) -N \gamma  \hat{f}(\beta,C) + N \gamma 
\hat{f}(\gamma,C)\cto,
\label{MIRA}
\eea
where  $\hat{f}(\beta,C)= f(\beta,C)-f(\infty,C)$ and
$f(\beta,C)$ is a function that is constant in each valley and it is  equal to the free 
 energy density
of the valley to which the configuration $C$ belongs.
In deriving eq.  (\ref{MIRA}) we have assumed that the value of $\gamma$ is 
high enough that all the configurations $C$, which contribute to the integral for large $N$, belong to 
some valley:  We have also assumed that there is a one to 
one correspondence among the valley at inverse temperatures $\beta$ and at $\gamma$.

Before entering into the computation of $\hat{f}(\beta,C)$ it is useful to make the so called quenched 
approximation, i.e. to make the following approximation inside the previous integral:
\be
\exp (-A \hat{f}(\beta,C))=\exp(-A \hat{f}_{\gamma}(\beta)), \label{APPROX}
\ee
where $\hat{f}_{\gamma}(\beta)$ is the expectation value of 
$\hat{f}(\beta,C)$ taken with the probability 
distribution proportional to $\exp(-\gamma H(C))$. The quenched approximation would be exact if the 
temperature dependance of the energy of all 
the valleys would be the same, apart from an overall shift at zero temperature. In other words we 
assume that the minima of the free energy have different values of the free energy but 
similar shapes. More refined computations can be done to compute systematically the 
corrections to the quenched approximation. This  approximation would be 
certainly bad if we were using the free energy $\hat{f}(\beta,C)$ at the place of ${f}(\beta,C)$ 
because the zero temperature energy strongly varies when we change the minimum.

We finally find 
\be
 \Phi(\gamma;\beta)= F_{L}(\gamma)+ \hat{f}_{\gamma}(\beta)  
-\hat{f}_\gamma(\gamma),
\ee
$F_{L}(\gamma)$ being the free energy of the liquid ($S_{L}(\gamma)$ will be the entropy of the 
liquid).
A simple algebra shows that 
\be
\Sigma(\gamma;\beta)=
S_{L}(\gamma)- S_{\gamma}(\gamma) +\hat{f}'_{\gamma}(\gamma) -
  \hat{f}'_{\gamma}(\beta), \label{FINAL}
\ee
where $
\hat{f}'_{\gamma}(\beta) = {\partial \hat{f}_{\gamma}(\beta) / \partial \gamma}
$.

In the liquid phase, we find out that the  configurational entropy is given by
\be
\Sigma(\beta)=\Sigma(\beta;\beta)=S_{L}(\beta)- S_{\beta}(\beta),
\ee
as  expected: the entropy of the liquid is  the entropy of the typical 
valley plus the configurational entropy.

The thermodynamic transition is characterized by the condition
\be
\Sigma(\beta_{K})=0.
\ee

In the glassy phase the free energy can be found by first computing the value of 
$\gamma(\beta)$ such that 
\be
\Sigma(\gamma(\beta);\beta)=0
\ee
and evaluating the corresponding free energy.

The quantity $\gamma(\beta)$ is the inverse of the effective temperature of the valley.  
It is easy to show  (following \cite{Mo}) that the previous formulae are completely equivalent to the 
replica approach. 

A strong simplification happens if we assume that the entropy of the valley can be evaluated in the 
harmonic approximation where we only keep the vibrational contributions.  In this case we obtain for 
a system with $M$ degrees of freedom that the harmonic entropy of the valley near to a 
configuration $C$ is given by
\be
S(\beta(C))=\frac{M}{2} \ln \ato {2 \pi e \over \beta}\cto -\frac12 \Tr\ato\ln (\cH(C) ) \cto,
\ee
where $\cH(C)$ is an $M \times M$ Hessian matrix (e.g. if $H$ depends on the coordinates $x_{i}$ 
we have that $\cH_{i,k}=\partial^{2} H / \partial x_{i} \partial  x_{k}$). 

If our approximation were fully consistent, we should find that all the eigenvalues of $\cH$ 
(the so called INN, Instantaneous Normal Modes \cite{Ke}) are positive.  This is not the case, 
however the number of negative eigenvalues becomes very small at low temperature, still in the 
liquid phase, signaling that valleys can be approximately defined in this region.

How can we evaluate the harmonic entropy in this case?  Two rather similar possibilities are: (a) We 
compute $\Tr\ato|\ln (\cH(C) )| \cto$ instead of $\Tr\ato\ln (\cH(C) ) \cto$.  (b) We find the 
minimum of the Hamiltonian which is the nearest to $C$ (i.e. the corresponding inherent structure) 
and we use the spectrum of the Hessian at this point.  We have checked numerically that the two 
methods give rather similar results \cite{LJ} and we will follow the second one in the numerical
simulations presented in the paper.

The framework has been set.  We now segue into an explicit computation.  Two possibilities are open: 
(a) We do an analytic computation of all the quantities which appear in the previous equations; (b) 
We extract them from numerical simulations.  The first possibility is open only in relatively simple 
systems and further approximations are needed, the second one is viable for all systems.

In these letter we shall explore both possibilities in the case of a {\sl realistic} model for 
glasses, the binary mixture of particles (80\% large particles, 20 \% smaller particles) interacting 
via a Lennard-Jones potential, introduced by Kob and Andersen \cite{KoAn}.  This Hamiltonian should 
mimick the behaviour of some metallic glasses and it is one of the best studied and simplest 
Hamiltonian which do not lead to crystalization at low temperature.

The numerical procedure does not present any serious difficulty.  We have studied via Monte Carlo 
simulations a system of $N=260$ particles, in a cubic box with periodic boundary conditions at 
density $\rho=1.2$, starting from $\beta^{1/4}=0.02$ and increasing $\beta$ by steps of $\Delta 
\beta^{1/4}=0.02$; we performed up to $4 \ 10^{6}$ MC steps at each $\beta$.  We need to simulate 
the system at high temperatures in order to compute the entropy of the liquid (we use as reference 
entropy that of a perfect gas at $\beta=0$).  The entropy is obtained using the formula 
$S(\beta)=S(0)+\int_{0}^{\beta}d\beta' (E(\beta)-E(\beta'))$.  Given an equilibrium configuration 
the near minimum of the potential is found by steepest descendent and the computation of the 780 
eigenvalues of $\cH(C)$ does not present any particular difficulty.

The results for the total entropy of the liquid and for the harmonic part are shown in fig.  (1) as 
function of $T^{-.4}$ (a more detailed description of the simulations can be found in ref.  
\cite{LJ}).  The entropy of the liquid is remarkable linear when plotted versus $T^{-.4}$, as it
happens in many cases \cite{Ro}.

In fig (2) we show the configurational entropy as function of $T^{-.4}$.  We fit it with a 
polynomial of second degree in $T^{-.4}$.  The extrapolated configurational entropy becomes zero at 
a temperature $T_{K}=.31\pm.04$, where the error contains systematic effects due to the 
extrapolation (similar conclusions have been reached in ref.  \cite{KST}).

\begin{figure}
\epsfig{figure=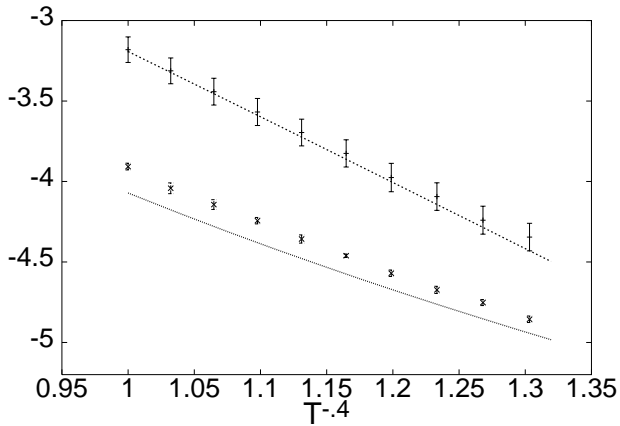,angle=270,width=8cm}
\vspace{.5cm}
\caption{ Analytical entropy of the liquid (upper line) compared with 
the numerical one, and analytical harmonic entropy 
(lower line) compared with the  numerical results. The horizontal axis is   
$T^{-.4}$.}
\end{figure}

There are many methods to compute analytically the free energy in the liquid phase which lead to 
integral equations for the correlation functions.  Here we follow a simple procedure proposed in 
ref. \cite{ZeHa}.  It mixes the HNC (hypernetted chain) and MSA (mean spherical 
approximation) closures by means of a single parameter $\al(T)$ that is chosen in order to reduce 
thermodynamic inconsistencies, minimizing the difference between two different ways of computing the 
compressibility. This technique allow us to compute the internal energy in the liquid with a 
reasonable approximation. The resulting integral equations for the correlations function of the two 
kinds of particles are transformed in a set on non linear differential equations by discretizing 
them with a space resolution of $2^{-5}$ and a large distance cutoff of $2^{4}$ (we have varied 
these parameters and checked that this choice gives a reasonable accuracy). The resulting equations 
in $3 \cdot 512$ unknown have been solved using a package from the IMSL library.

The computation of the spectrum is much more involved.  We  follow a simple approach which 
becomes exact if we assume that the range of the 
Hessian is much larger than the typical interatomic distance and we use the superposition 
approximation
\be
g(x_{1}x_{2}x_{3})=g(x_{1}-x_{2})g(x_{2}-x_{3})g(x_{3}-x_{1})
\ee
in the cases where it is needed \cite{MePa1,sferesoft,LJ}.
In order to compute the eigenvalues of the Hessian we compute analytically the moments of the 
eigenvalue distribution
\be M_{k}=\Tr (\cH ^{k}).
\ee 
Each moment can be written as the appropriate integral over the correlations functions.  In this long 
range approximation some terms are dominant over the others.  We keep only those terms and we 
introduce the superposition approximation into the appropriate places.  We use the large range 
expansion to select the diagrams \cite{MEPAZEE}.  

After some computations (which we do not report here \cite{MePa1,sferesoft,LJ}) we find a 
spectral density $\rho(e)$ that has support only in the region of positive eigenvalues $e$ (real 
frequencies $\omega=\sqrt e$) and goes to zero as $\omega$ as expected.
\begin{figure}
\epsfig{file=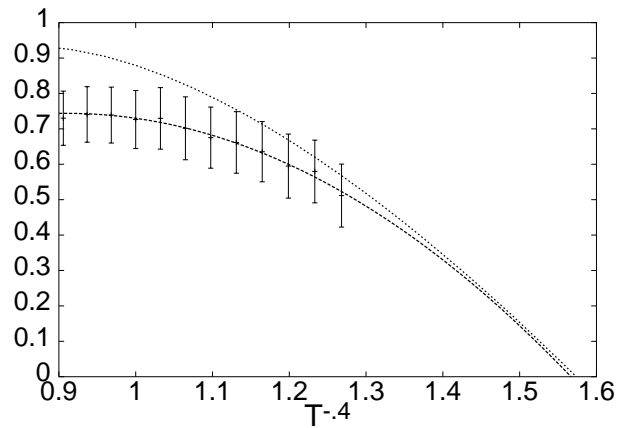,angle=270,width=8cm}
\vspace{.5cm}
\caption{ Analytical value of the complexity (upper line) compared with 
the numerical one ($+$ points), as functions of $\beta^{.4}$.}
\end{figure}

We can now put everything together: the final analytic predictions for the liquid and harmonic 
entropy are shown in fig.  (1).  The liquid entropy turns out to be very good, while there is a 
minor discrepancy for the harmonic entropy, likely due to the rather strong approximations we have 
done in the analytic computation.  The analytic configurational entropy is shown in fig.  (2).  It 
becomes zero at $T_{K}=.32$, which is our analytic prediction for the thermodynamic transition (as a 
check we have fitted the analytic configurational entropy using the same procedure as for the 
numerical one and we have found $T_{K}=.34$).

We have in our hands all the tools to compute analytically the free energy in the low temperature 
case.  In fig.  (3) we show the specific heat coming from the $x$ dependent part of the Hamiltonian 
as a function of the temperature (we must add the momentum contribution $3/2$ in order to get the 
total specific heat).  Very similar results are obtained by using the extrapolated numerical 
entropies at the place of the analytic ones.  We notice that the Dulong Petit law is extremely well 
satisfied in the low temperature region.  Indeed in the harmonic approximation the Dulong Petit law 
would be exact if we neglect the $\gamma$ dependence of $S_{\gamma}(\beta)$.  The value of 
$\gamma(\beta)$ weakly depends on $\beta$: its value in the limit $\beta \to \infty$ is only about 
10\% higher that its value (i.e. $\beta_{K}$) at the transition temperature.

The results presented here are an explicit realization of a transition driven by an entropy crisis, 
which has been firstly implemented microscopically in the random energy model \cite{De}.  The 
specific heat has a jump downward, when we decrease the temperature, which is the opposite of the 
typical behaviour in transitions characterized by the onset of a conventional order parameter (in 
mean field theory ferromagnets, superconductors, have a jump upward when we decrease the 
temperature).

\begin{figure}
\epsfig{figure=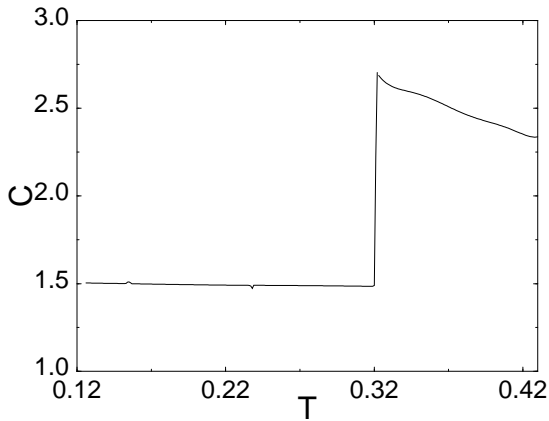,angle=270,width=8cm}
\vspace{.5 cm}
\caption{The specific heath coming from the $x$ dependent part 
of the Hamiltonian as a function of the temperature}

\end{figure}

We stress that it is possible to remove the approximation of using the harmonic entropy.  There are 
no serious difficulties in computing numerically the true entropy of the valleys; this has been done 
for a binary mixture of soft sphere and the results are very near to the harmonic ones \cite{sferesoft}.  
Analytically we can also expand in the anharmonicity parameters.  It is also  possible to take 
care of the fluctuations of the entropy from valley to valley and go beyond the quenched 
approximation.  It is quite likely that these effects will not strongly change the results.  The 
most important step would be to reach a better theoretical control on the spectrum of the INN in the 
liquid phase, maybe combining the approximations used here with the low density expansion of 
\cite{CAGIAPA}.

Summarizing we have found a method which is able to use liquid theory method in the glasses phase 
putting in practice the old adage {\sl a glass is a frozen liquid}.  We are able to compute with a 
reasonable approximation the thermodynamics and with a little more effort we can compute the static 
and the dynamic structure functions.

We owe a lot to M.M\'ezard and we are very happy to thanks him.  We are also very grateful to W.Kob 
and F.Sciortino for their suggestions and to A.Cavagna and I.Giardina for useful discussions.  B.C. 
would like to thank the Physics Department of Rome University `La Sapienza` where this work was 
partially developed during her PhD.


\begin{thebibliography}{99}

\bi{MePa1} M. M\'ezard and G. Parisi, {\em Phys.  Rev.  Lett.}  
{\bf 82} 747 (1998); 
M. M\'ezard and G. Parisi, {\em J. Chem. Phys.} {\bf 111 No.3}, 1076 (1999).

\bi{Me} M. M\'ezard, {\em How to compute the thermodynamics of a glass using a cloned liquid}, 
cond-mat/9812024.

\bi{sferesoft} B. Coluzzi, M. M\'ezard, G. Parisi and P. Verrocchio, 
{\em J. Chem. Phys.} {\bf 111 No.19},9039 (1999)

\bi{LJ} B. Coluzzi, G. Parisi and P. Verrocchio, {\em Lennard-Jones
  binary mixture: a 
thermodynamical approach to glass transition}, cond-mat/9904124.

\bi{KiThWo} T.R. Kirkpatrick and P.G. Wolynes, {\em Phys.  Rev.} {\bf A 34}, 1045 (1986); T.R. 
Kirkpatrick and D. Thirumalai, {\em Transp.  Theor.  Stat.  Phys.} {\bf 24}, 927 (1995) and 
references therein.

\bi{Mo} R. Monasson, {\em Phys. Rev. Lett.} {\bf 75}, 2847 (1995).

\bi{GoSj} 
For a review see W. G\"otze and L. Sj\"ogren, {\em Rep. Prog. Phys.} {\bf 55}, 241
(1992).

\bi{SaDeSt} S. Sastry, P. G. Debenedetti and F. H. Stillinger, {\em Nature} {\bf 393} 554 (1998).

\bi{landscape} For example see also F. Sciortino and P. Tartaglia, {\em Phys. Rev. Lett.}
{\bf 78}, 2385 (1997).  K. K. Bhattacharya, K. Broderix, R. Kree, A. Zippelius, cond-mat/9903120.  L. Angelani, G. 
Parisi, G. Ruocco and G. Viliani, {\em Phys. Rev. Lett.} {\bf 81} 4648 (1998)

\bi{Ka}A.W. Kauzman, {\em Chem. Rev.} {\bf 43}, 219 (1948).

\bi{An} For a recent review see C.A. Angell, {\em Science} {\bf 267}, 1924 (1995) and P. De 
Benedetti, {\em Metastable liquids}, Princeton University Press (1997).

\bi{St} F.H. Stillinger and T.A. Weber, {\em Phys.  Rev.} {\bf A 25}, 2408 (1982).  F.H. Stillinger, 
{\em Science} {\bf 267}, 1935 (1995) and references therein.

\bi{KST1} W. Kob, F. Sciortino, P. Tartaglia {\em Aging as dynamics in configuration space},
cond-mat/9905090.

\bi{Ke} See for instance T. Keyes, {\em J. Phys. Chem.} A {\bf 101},
2921 (1997), and reference therein.


\bi{KoAn} W. Kob and H.C. Andersen, {\em Phys.  Rev.  Lett.} {\bf 73}, 1376 (1994).  

\bi{Ro} Y. Rosenfeld and P. Tarazona, {\em Molecular Physics} {\bf
  95}, 141 (1998).
  
\bi{KST} W. Kob, F. Sciortino, P. Tartaglia, {\em Inherent Structure Entropy of Supercooled Liquids},
cond-mat/9906081.

\bibitem{ZeHa}
G.Zerah and J.P. Hansen, {\em J. Chem. Phys.} {\bf 84} (4), 2336 (1986).


\bi{MEPAZEE} M. M\'ezard, G. Parisi and A. Zee {\em Spectra of Euclidean Random Matrices}, 
cond-mat/9906135.

\bi{De} B. Derrida  {\em Phys. Rev.} {\bf B24}  2613 (1981).

\bi{CAGIAPA} A. Cavagna, I. Giardina and G. Parisi, 
{\em Phys.  Rev.  Lett.} {\bf 83}, 108 (1999).

\end{thebibliography}
\end{document}